\documentclass[prl,superscriptaddress,showpacs,twocolumn]{revtex4-1}
\usepackage{amsmath,graphicx,amssymb,amsfonts}%,times
\begin{document} \title{Why Granular Media Are, After All,  Thermal}
\author{Yimin Jiang} 
\affiliation{Central South University, Changsha 410083, China}
\author{Mario Liu}
\affiliation{Theoretische Physik, Universit\"{a}t T\"{u}bingen,72076
T\"{u}bingen, Germany}
\date{\today}

\begin{abstract} 
Granular media are considered {\it athermal}, because the grains are too large to display Brownian type thermal
fluctuations. On the other hand, being macroscopic, every grain undergoes thermal expansion, possesses a temperature
that may be measured with a thermometer, and consists of many, many { internal degrees of freedom that in their sum do
affect granular dynamics.} Therefore, including them in a comprehensive approach to account for granular behavior
entails crucial advantages. The pros and cons of thermal versus athermal descriptions are considered.
\end{abstract}

\maketitle 
%%%%%%%%%%%%%%%%%%%%%%%

Taking grains as athermal particles interacting via the Newtonian law with an elasto-frictional force, discrete element
method (DEM) has been a success story~\cite{dem1,dem2,dem3}, to the extent that it is today the tool of choice for
coming to terms with granular behavior~\cite{dem4}. A second success story is the granular kinetic
theory~\cite{kin0,kin1,kin2,kin3,kin4} that also takes grains as athermal paticles colliding with a restitution
coefficient smaller than one. As a result -- and because granular Brownian motion is imperceptibly small -- it is a
common place of the granular community that grains may generally be approximated as athermal particles.

This believe is best reflected in athermal statistical mechanics (ASM) that defines, in addition, an entropy $S$ (as the
logarithm of the number grains may be stably packed), assuming it is maximal in equilibrium, if the numbers are
appropriately weighted~\cite{Edw,raphi,1,1nico,2}. (In contrast, DEM neither needs an entropy to guide the system
towards the rest state, nor ever introduces it.) Being nonconserved, energy is typically discarded as a state variable.
Instead, $S$ is taken to depend on the volume and force moment tensor. Force equilibrium is often assumed.

This is a leap of faith: 
Granular Brownian motion is small because grains are macroscopic. But there are then many, many internal microscopic
degrees of freedom that one needs to make sure are irrelevant. To see whether they are, consider the textbook
example of a pendulum. Its motion is given by the Newtonian force law of a mass point, including a friction term -- no
need to consider any microscopic degrees here. But to determine the sign of the friction, to make sure that the
amplitude diminishes, one needs to consider how the total entropy, consisting mainly of microscopic degrees of freedom,
increases. These are the air molecules surrounding the pendulum and the phonons in the string -- also the phonons and
electrons in the pendular weights, if there are two steel pendula colliding periodically. When the pendula come to a
standstill hanging down, their macroscopic energy is zero, and {\it the total entropy maximal.}

If this is an apt analogy for the relation between DEM and granular statistical mechanics, if the microscopic degrees of
freedom do influence the macroscopic dynamics, they must be included when calculating the entropy.

At any rate, if we consider the air or water surrounding the grains, we need to include their molecular degrees of
freedom in an entropic considerations: Is it not consistent to include the inner-granular ones as well?

Doing so, helpfully, the total energy is conserved, and the thermal statistical mechanics valid -- as is thermodynamics.
Employing them, one finds that the grain-level energy dissipates because it is being redistributed to the microscopic
degrees of freedom, and that force equilibrium holds as a result of $S$ being maximal. 

We therefore submit: {\it Granular media are not generally athermal. More specifically, taking a reduced entropy that
excludes internal degrees of freedom as maximal in equilibrium, one bears a heavy burden of proof.}\\

\begin{figure}[t] 
\begin{center}
\vspace{-2.5cm}
\includegraphics[scale=0.2]{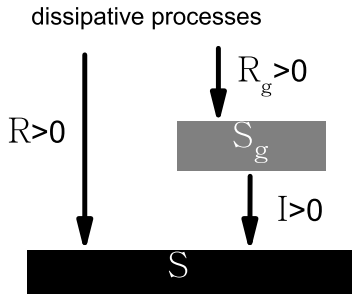}
\end{center}
\vspace{-2cm}
\caption{\label{2stageIrr}
Dissipative processes produce either granular entropy $S_g$, or thermal entropy $S$. Eventually, $S_g$ also decays to
$S$. Irreversibility is implied by the three energy-decay terms $R$, $R_g$, $I$ being always positive. Containing all states of deformed and moving grains, $S_g$ is a much larger quantity than various versions of the reduced entropy in ASM, though $S_g\ll S$ still holds. The system is in
equilibrium if $S_{tot}\equiv S+S_g\approx S$ is maximal. 
Maximal $S_g$ characterizes (quasi-)equilibrium for times $t\ll\tau$ (with $\tau$ the time scale of $I$). Then
conventional statistical mechanics and thermodynamics hold. Maximal $S_g$ for $t\to\infty$ is a novel proposition. }
\end{figure}

\noindent{\bf Two-stage irreversibility}, a useful notion for coming to terms with granular thermodynamics is related to
the three spatial scales of any
granular media: (a)~the macroscopic, (b)~the mesoscopic or inter-granular, and (c)~the
microscopic or inner granular. Dividing all degrees of freedom into these three
categories, one treats those of (a) differently from (b,c). Macroscopic
degrees of freedom (the slowly varying stress, flow and density fields) are
employed as state variables, but inter- and
inner granular degrees of freedom  are treated summarily: 
Only their contributions to the energy is considered and taken,
respectively, as granular and true heat. One does not account for the microscopic dynamics of phonons and electrons, but
takes the sum of their energy as $\int T{\rm d}S$. Similarly, one does not account for the motion and deformation of
every grain, only includes their fluctuating kinetic and elastic energy as granular heat $W_T$. Defining $S_g$ as the
logarithms of the number of states the inter-granular degrees may be in, and
$T_g\equiv\partial W_T/\partial S_g$,
we again have  $W_T=\int T_g{\rm d}S_g$. 
There are a handful of macroscopic degrees of freedom~(a), a large number of inter-granular ones~(b), and yet many orders of magnitude more inner granular ones~(c). So the statistical tendency to equally distribute the energy among all
degrees of freedom implies an energy decay: (a $\to$ c) and (a $\to$ b $\to$ c), or what we termed {\em two-stage irreversibility}, see
Fig~\ref{2stageIrr}.

Starting from two-stage irreversibility and the resultant granular thermodynamics, a hydrodynamic theory for granular
media (named GSH for ``granular solid hydrodynamics") was derived~\cite{granR2,granR3,granR4,granRgudehus}, and it has
proven capable of accounting for a wide range of granular phenomena, including (i)~static stress distribution, clogging;
(ii)~elasto-plastic motion: loading and unloading, approach to the critical state, angle of stability and repose;
(iii)~rapid dense flow: the $\mu$-rheology, Bagnold scaling and the stress minimum; (iv)~elastic waves, compaction, wide
and narrow shear band; (v) the less conventional experiments of shear jamming, creep flow, visco-elastic behavior and
nonlocal fluidization~\cite{granRexp}.

A division into three scales works well when they are clearly separated -- though this is a problem of accuracy, not
viability. Scale separation is well satisfied in large-scaled, engineering-type experiments, but less so in small-scaled
ones. Using glass or steel beads aggravates the problem. The same is true of 2D experiments employing less and larger
disks. On the other hand, when there is too little space for spatial averaging, one may still average over time and
runs, to get rid of fluctuations not contained in a hydrodyanmic theory.
\\

\noindent{\bf The fluctuation-dissipation theorem (FDT)} 
correlates the thermal fluctuations of the microscopic degrees of freedom in equilibrium to how they return to
equilibrium after having been exposed to a small applied perturbation, ie., to the friction and dissipation that they
experience~\cite{kubo}. The microscopic degrees of freedom are atoms or molecules in gas and liquid, and phonons, free
electrons in solid. So there is no reason why FDT, a general principle, should not be valid in granular media: FDT holds
in a block of copper, although it is much larger than a grain and has even smaller Brownian motion.
This remains so when one cuts the block in half, because a cut is too small and macroscopic a perturbation to
appreciably affect FDT, or the linear response theory employed to deduce it. Further cuts are also allowed, as long as
the pieces remain macroscopic. The link between thermal fluctuations and the return to equilibrium will not be severed
-- although the link may change, because the system's hydrodynamics (such as given by GSH) does. This conclusion clearly holds for all
granular media.

The alleged problem with FDT lies elsewhere: Due to random packaging, grains as such also fluctuate -- with respect to
their position, velocity and external forces. Although these grain-level, non-thermal fluctuations are much stronger and 
the ones being observed, there is, {\it prima facie}, no general principle linking them to dissipation. Still, 
attempts abound to find a $T$ from these fluctuations such that FDT does hold, see~\cite{1,2,hou}.
There are many observations on whether FDT is valid, experimental and numerical, and the considered link is always to
grain-level fluctuations: Thermal fluctuations are universally set to zero in DEM, and it is difficult, if at all
possible, to separate both in experiments. \\

\noindent{\bf The Onsager relation (OR)} is a direct consequence of FDT. If the latter holds, so does the former. But
there is also an independent counter-argument that we need to deal with: The validity of the OR depends on the {\it time
reversal invariance} of the underlying microscopic dynamics, yet grains collide inelastically and execute irreversible
motion. This is why many believe OR need not hold in granular media, see~\cite{onsager}. This point of view again ignores the fact
that grains are not elementary. The granular kinetic theory is a truncated one, and the apparent lack of time reversal
symmetry is a result of this truncation. The microscopic degrees of freedom in sand obey, as everywhere else, the {\it
reversible} Schr\"{o}dinger equation.

OR concerns the pairwise equality (with same or opposite signs) of transport coefficients, a fact that may be verified
experimentally. Although no one is capable of employing the Schr\"{o}dinger equation to calculate the coefficients
directly, there should be little doubt that the result will comply with the OR. Therefore, a proper calculation
employing the dissipative kinetic theory must arrive at the same result. (This may not be easy, as the kinetic theory at
present only considers $R_g$, not $R$, see Fig 1.) More generally,
OR is valid in all condensed systems: solids, superfluids, liquid crystals, because all share the same microscopic
dynamics. Breaking a solid block into (macroscopic) pieces does not change this fact.\\

\noindent{\bf Is $\mathbf{T_{\pmb g}}$ a true temperature?} Yes, $T_g$ and $T$ each characterizes the energy of a group
of degrees of freedom, with a given rate of energy transfer between them. Usually, the degrees of freedom of two systems
in thermal contact are spatially separated. In this case, they are not, which is the main difference. Any
difficulties treating $T_g$ as a temperature arise because one ignores $T$ -- as in the cases discussed above, and it is
also the reason for the failure of granular media to equilibrate, that granular temperatures of two systems in contact
are different. 

Given two granular systems, 1 and 2, with only 1 being excited, there are, in the steady state, four generally unequal
temperatures: $T^1,T_g^1,T^2,T_g^2$, and an ongoing energy transfer $(T^1_g\to T^1)$, $(T^1_g\to T_g^2)$ and $(T_g^2\to T^2)$. Their
differences depend on details such as the restitution coefficients and the contact area.
This is quite similar to four thermal systems, two large (1, 2), two small (1a, 2a), with (1a) being heated, and (1a,
1), (1a, 2a), (2a, 2) in pairwise thermal contact. All four temperatures will usually be different,
and there is an ongoing energy flux.
 
The passive granular system 2 may serve as a thermometer if it consisted of completely elastic beads, and if both are
separated by a massless membrane that transmits momentum but no particles. (So this is more a DEM-thermometer.) In
the steady state, we have $T^1_g=T_g^2$, and the energy transfer $(T^1_g\to T_g^2)$ vanishes.
Similarly, if system 2 does not exists, we have $T_{a}^1=T_{a}^2$.

Finally, a caveat: Employing $T_g$ to quantify the granular degrees of freedom is sensible only if these are in {\it equilibrium with one another}. This is not always the case, eg. a granular gas maintained by vibrating walls, which therefore needs additional state variables, see~\cite{GGas}. But grains become increasingly better equilibrated for higher densities and longer lasting contacts. To keep the discussion simple, we assume here that they are. \\

\noindent{\bf Granular Thermodynamics} has, as its set of state variables, first the usual ones: the density $\rho$, the
momentum density $\rho v_i$, the total entropy $s_{tot}$; then in addition: the granular entropy $s_g$, the elastic
strain $u_{ij}$. (The elastic strain $u_{ij}$ is a coarse-grained measure of the grains' elastic deformation, and not
the total strain, see~\cite{granR2,granR3,granR4}.) Denoting the
conserved energy density (in the rest frame, $v_i=0$) as $w=w(s, s_g, \rho, u_{ij})$,
the conjugate variables are: $T\equiv{\partial w}/{\partial s}$, $T_g\equiv{\partial
w}/{\partial s_g}$, $\mu\equiv{\partial w}/{\partial\rho}$,  $\pi_{ij}\equiv-{\partial w}/{\partial u_{ij}}$, 
where $\mu$ is the  chemical potential and $\pi_{ij}$ the elastic stress. This is conveniently written as
\begin{equation}\label{2-1}
{\rm d}w=T{\rm d}s+T_g{\rm d}s_g+\mu{\rm d}\rho-\pi_{ij}{\rm d}u_{ij},
\end{equation} 
a formula valid as long as the variables are as given. Writing
$T{\rm d}s+T_g{\rm d}s_g=T{\rm d}(s+s_g)+(T_g-T){\rm d}s_g$,
we identify the first term as the equilibrium energy for $T_g=T$, and the second as the additional contribution if
$T_g\not=T$. 
Characterizing the non-optimal energy distribution between the inter- and inner granular degrees of freedom, $T_g-T$ relaxes until it vanishes.
Now, since $s\gg s_g$, and since any granular motion at all already implies  $T_g\gg T$, we have $T_g-T\approx T_g$, 
$s+s_g\approx s$, and this rewriting does not change anything. So we may  take $T_g$ as 
the relaxing quantity, with $T_g=0$ replacing $T_g=T$ at equilibrium. [This result is essential for obtaining
Eq.(\ref{eq1}).]

Formal equilibrium conditions, given in terms of
the conjugate variables and valid irrespective what the actual expression for $w=w(s, s_g, \rho, u_{ij})$ is, are
obtained
by requiring $\int s\,{\rm d}^3r$ 
to be maximal, under the constraints of constant energy $\int w\,{\rm d}^3r$ and mass $\int \rho\, {\rm d}^3r$. In
granular media, remarkably, this universally
valid procedure leads to two sets of equilibrium conditions, 
solid- and fluid-like. 

Maximizing the entropy (see~\cite{granR2,granR3,granR4} for details), we always 
obtain $\nabla_iT=0$, and $T_g=0$. Usually, $T_g$
vanishes quickly, and if it does, the density no longer varies independently from
the elastic strain, $d\rho/\rho=-du_{\ell\ell}$. They then share a
common,  solid-like equilibrium condition,
\begin{eqnarray}
\label{2a-1}
\nabla_i(\pi_{ij}+P_T\delta_{ij})=\rho\, {\rm g}_i,
\end{eqnarray} 
where ${\rm g}_i$ is the gravitational constant, $\pi_{ij}$ the elastic stress,
$P_T\equiv-\partial(wV)/\partial V$ the usual fluid pressure ($V$ is the volume, and the derivative
is taken at constant $\rho V$, $s_{tot}V$ and $s_gV$).
With Eq.(\ref{eq1}) below, $P_T\sim T_g^2\to0$ is the pressure exerted by jiggling grains. Clearly, Eq~(\ref{2a-1}) expresses force balance in the jammed state.

If $T_g$ is kept finite by continual external perturbations, the system further increases its entropy by independently
varying $\rho$ and $u_{ij}$, to arrive at the fluid equilibrium, characterized by two conditions. The first, with
respect to $u_{ij}$,
requires any shear stress to vanish and any free surface to be horizontal in equilibrium; the second is related  
to reversible compaction, see~\cite{granRexp} for more details: 
\begin{equation}\label{2a-2} \pi_{ij}=0, \quad
\nabla_iP_T=\rho\, {\rm g}_i. \end{equation} 

\noindent{\bf The relation between $\mathbf T_{\pmb g}$ and $\mathbf S_{\pmb g}$:}
In the gaseous phase, grains have only kinetic energy, $\frac12T_g$ per degree of freedom. With $N$ the number of
grains, the total energy is $W_T=\frac32T_gN$.
If the inner granular degrees of freedom may be modeled as a phonon gas, the inner energy is $3TN_a$ (for $T\gg T_D$, the Debye temperature, and with $N_a$ the
number of atoms in all the grains). Assuming (unrealistically) that the
grains maintained their integrity at arbitrarily high $T$, they will heat up during a collision for $T_g>T$, but cool
down for $T_g<T$, until $T_g=T$, a result associated with the total energy being
conserved. Usually, of course, because $T_g\gg T$, the heat transfer is accounted for by a constant restitution coefficient. 
 
This picture becomes blurred at higher densities,  breaking down completely when the contacts are enduring. 
Given the friction between grains, a suitable $w(s_g)$ is difficult to obtain microscopically. Therefore, we
pragmatically
expand $w$ in $s_g$, requiring $T_g=0$ to be minimal. Denoting $w_T=w-w(T_g=0)$, we have
\begin{equation}\label{eq1}
w_T={s_g^2}/{2\rho b}=\rho b{T_g^2}/2,\quad T_g\equiv{\partial w}/{\partial s_g}|_{s}={s_g}/{\rho b}.
\end{equation}
As this expression assumes only analyticity of $w$ and does not depend on interactions, it is quite general -- as long
as $T_g$ is sufficiently small.

One could now find an appropriate formula interpolating between this limit and the gaseous one. But we, rather more
simply, employ
Eq.(\ref{eq1}) for all values of $T_g$. This works because of $T_g$'s two functions, as a measure for equilibration and
as a state variable, only the second is elevant in praxise: Substituting $w_T=$ $\rho b{T_g^2}/2$ for
$w_T=\frac32T_k\rho/m$ implies $T_k=\frac13mb T_g^2$ in the gaseous limit ($m$ is the average mass of a grain, 
$T_k$ is introduced to distinguish both). This is impermissible
when one considers equilibration, as done above, because it is $T_k$ that becomes equal to $T$, not $T_g$. On the other
hand, such super hot grains do not exist, and equilibration (though helpful for coming to terms with $T_g$) is not a
realistic scenario.
The second, and relevant, role of $T_g$ does not possess such a scale.
For instance, the pressure is found $\sim T_k$, the viscosity $\sim\sqrt{T_k}$, in the kinetic theory and in DEM,
see~\cite{luding2009,Bocquet}, while they are, respectively, $\sim T_g^2$ and $\sim T_g$ in GSH. So Eq.(\ref{eq1}) may
be taken as valid throughout, and in fact defines $T_g$. \\

\noindent{\bf Conclusions:}  
Given the three length scales in any granular media: macroscopic, granular and microscopic,  two temperatures: $T, T_g$ are relevant, and the entropy is the sum of all granular and microscopic degrees of freedom.
Thermodynamics holds, and $T_g$ is well-behaved, if one includes $T$ in all considerations. A jammed state at rest is in equilibrium, with the entropy being maximal. Fluctuation-dissipation theorem, correlating dissipation to thermal fluctuations, holds in granular media as in any other system. 
The strong, nonthermal  grain-level stochasticity -- such as force chains, or stick and slip motion --  has yet to find a theory. But the macroscopic granular dynamics, averaged over
thermal and grain-level fluctuations, is well accounted for by GSH.

\end{document}